\documentclass[amsmath,prb,twocolumn,showpacs,superscriptaddress]{revtex4}

\usepackage{color}
\usepackage{amsfonts}
\usepackage{graphicx}

\begin{document}

\title{Transport in molecular states language:\\ Generalized quantum master equation approach}
\date{\today}
\author{Massimiliano Esposito}
\altaffiliation[Also at]{Center for Nonlinear Phenomena and Complex Systems,
Universit\'e Libre de Bruxelles, Code Postal 231, Campus Plaine, B-1050 Brussels, Belgium}
\author{Michael Galperin}
\affiliation{Department of Chemistry \& Biochemistry, University of California San Diego, La Jolla CA 92093, USA}

\begin{abstract}
A simple scheme capable of treating transport in molecular junctions
in the language of many-body states is presented. An ansatz in Liouville 
space similar to generalized Kadanoff-Baym approximation is introduced
in order to reduce exact equation-of-motion for Hubbard operator to
quantum master equation (QME)-like expression. A dressing with effective 
Liouville space propagation similar to standard diagrammatic dressing
approach is proposed. The scheme is compared to standard QME approach,
and its applicability to transport calculations is discussed within 
numerical examples.   
\end{abstract}


\maketitle
\section{Introduction}\label{intro}

Quantum transport in nanoscale systems is on the forefront of research in 
many fields of study. In particular, progress of experimental capabilities 
in the field of molecular electronics brings new theoretical 
challenges.\cite{science_review}
Resonant transport with strong on-the-bridge interactions is one of them.
It is probably the most important regime for possible future applications 
(e.g. for logic and memory molecular devices). Unlike usual mesoscopic systems,
molecular electronic (and vibrational) structure may be very sensitive to
reduction/oxidation. Thus resonant transport in molecular junctions has to
be described in the language of  molecular states (states of the isolated
molecule as a basis for description of reduced non-equilibrium dynamics),
rather than in the language of (effective) single-particle orbitals,
which is generally accepted in the molecular electronics community.
Recent experiments on simultaneous measurements of current and optical 
response of molecular junctions\cite{natelson} make 
need for such formulation even more pronounced, since molecular states is
a natural language of all the (equilibrium) molecular spectroscopy.
Besides, molecular states based formulation of transport makes it 
potentially possible to incorporate standard quantum chemistry
molecular structure simulations as an input to transport calculations.

Necessity of many-body states type description of transport in molecular
systems has been realized. Among the approaches one can mention
scattering theory,\cite{BoncaTrugman} 
(generalized) master equation,\cite{Datta,Datta_ph,Koch2005,Koch2006,Koch2006a,Koch2007}
density matrix,\cite{Petrov2005,Petrov2006,Schoeller,Wacker05,Yan05,Schreiber06,HarbolaEsposito06}
and non-equilibrium Green function (NEGF) based schemes.\cite{Sandalov,Fransson,hubop}
Each of these schemes has its own limitations. Scattering theory 
when applied to transport problems disregards junction character
of molecular system which may lead to erroneous 
predictions.\cite{MitraAleinerMillis,strong_elph}
Standard formulations also miss crucial physics, e.g. effective attractive
electron-electron interaction via phonons (bipolaron formation),
energy exchange (heating and cooling effects) between successive
tunneling events, target distortion due to quasi-bound states etc.
Master equation (or generalized master equation) approaches are  
used quite often to describe hopping transport, i.e. situation when
correlations in the system (both in space and time) 
die much quicker than electron transfer time (fixed by 
contact/molecule coupling). Besides, they become
inadequate in the off-resonant tunneling (super-exchange) situation. 
Density matrix schemes usually formulated within quantum master equation 
form, often miss broadening of molecular states due to coupling to 
the contacts and coherences responsible, e.g., for elastic channel 
renormalization at the inelastic threshold (see also discussion below).
NEGF based approaches in the language of molecular states are among
the most advanced methods for treating non-equilibrium molecular 
systems.\cite{hubop} Two important drawbacks of the approach are its
complicated character and absence of proper commutation relations
for the Hubbard operators. The first means that applicability of
the method is limited to simple cases only. The second may lead to
unphysical consequences (e.g. non-Hermiticity of the reduced density
matrix) at approximate level of treatment.\cite{Novotny,Novotny2}
  
The goal of the present paper is to formulate an approximate scheme
for treatment of transport in molecular junctions in the language
of molecular states, exploring connection between Green function and 
density matrix based approaches to transport.
We start from the NEGF-like consideration
and derive QME-like equation, pointing out approximations involved in
the derivation. Note that similar approach within a single-particle orbitals 
language was used in Ref.~\onlinecite{Neuhauser}.
Note also that in our consideration we go beyond strictly Markov limit 
of Ref.~\onlinecite{Wacker05}.
We work with Hubbard operator as a natural
object capable of describing excitation in the molecule as transitions
(we restrict our consideration to single-electron transitions)
between many-body molecular states. First we show how exact equation
can be reduced to QME by introducing ansatz in Liouville space 
similar to the generalized Kadanoff-Baym approximation (GKBA)\cite{HaugJauho} 
in Hilbert space. Second, we identify diagrams on the Keldysh contour 
corresponding to the processes described by QME, and in the spirit of
Green function diagrammatic techniques dress this diagrams by effective
Liouville space dynamics. The last is obtained from the exact equation 
within Markov approximation. Section~\ref{model} introduces molecular
junction model and presents equation-of-motion for the Hubbard operator. 
In section~\ref{qme} we introduce an ansatz in the Liouville space, which
being similar to the GKBA, allows reduction of the exact EOM for the Hubbard
operator to QME. Here we also discuss a dressing procedure.
Section~\ref{results} presents analytical consideration for the simple
resonant level model, and discuss general numerical procedure. 
Several numerical examples are presented as well. 
Section~\ref{conclude} concludes.

\section{\label{model}Model}

We consider molecular junction which consists of two contacts ($L$ and $R$)
coupled through the molecule ($M$). Contacts are assumed to be reservoirs 
of free electrons each at its own equilibrium. All the non-equilibrium 
physics takes place at the molecule. Hamiltonian of the system is
\begin{equation}
 \label{Htot}
 \hat H = \hat H_L + \hat H_M + \hat H_R + \hat V 
 \equiv \hat H_0 + \hat V
\end{equation}
where $\hat H_M$ is a full molecular Hamiltonian, i.e. Hamiltonian
of isolated molecule with all on-the-molecule interactions included,
$\hat V$ is molecule-contacts coupling 
\begin{equation}
 \label{V}
 \hat V = \sum_{m\in M, k\in\{L,R\}}\left(
 V_{km} \hat c_k^\dagger\hat d_m + V_{mk} \hat d_m^\dagger\hat c_k
 \right)
\end{equation}
and 
$\hat H_K$ ($K=L,R$) represents contacts
\begin{equation}
 \label{HK}
 \hat H_K = \sum_{k\in K} \varepsilon_k \hat c_k^\dagger\hat c_k
\end{equation}
Here $c_k^\dagger$ ($\hat c_k$)  and $d_m^\dagger$ ($\hat d_m$) 
are creation (annihilation) operators for electron in a state $k$ 
in contact $K$ and state $m$ of molecular Hamiltonian, respectively.

We introduce many-body states of isolated molecule $|N,i>$ with
$N$ being number of electrons on the molecule and $i$ standing for a set 
of all other quantum numbers characterizing particular state of the molecule
in the charging block $N$. These states are assumed to be orthonormal
\begin{equation}
 \label{orthonormal}
 <N,j|N',i'> = \delta_{N,N'}\,\delta_{i,i'}
\end{equation}  
Note, that generalization to nonorthogonal basis is available in the 
literature,\cite{Sandalov_ovlp} but we will stick with orthonormal basis 
in order to keep notation as simple as possible. Molecular transitions
(in our case due to coupling to the contacts) are naturally described in the
language of Hubbard operators
\begin{equation}
 \label{X}
 \hat X_{(N,i;N',i')} = |N,i><N',i'|
\end{equation}
One of important (in our case) transitions is oxidation/reduction
of the molecule by one electron, i.e. transition between neighboring
charge blocks
\begin{equation}
 \mathcal{M} \equiv (N,i;N+1,j) \qquad
 \bar{\mathcal{M}}\equiv (N+1,j;N,i)
\end{equation}
In terms of these states molecular Hamiltonian is
\begin{equation}
 \label{HM}
 \hat H_M = \sum_{N,i,j} |N,i> H_{ij}^{(N)} <N,j|
 \equiv \sum_{N,i,j} H_{ij}^{(N)} \hat X_{(N,i;N,j)}
\end{equation}
If the many-body states are chosen to be eigenstates of the molecular
Hamiltonian, $H_{ij}^{(N)} = E^{(N)}_i \delta_{i,j}$ with
$E^{(N)}_i$ being energy of the molecular eigenstate $|N,i>$. 
Molecule-contacts couping, Eq.(\ref{V}), becomes
\begin{equation}
 \label{VMB}
 \hat V = \sum_{k,\mathcal{M}} \left(
   V_{k\mathcal{M}} \hat c_k^\dagger \hat X_\mathcal{M}
 + V_{\bar{\mathcal{M}}k} \hat X_{\bar{\mathcal{M}}}\hat c_k
 \right)
\end{equation}
where 
\begin{equation}
 \label{VME}
 V_{k\mathcal{M}} \equiv \sum_{m\in M} V_{km} <N,i|\hat d_m|N+1,j>
\end{equation}
and $V_{\bar{\mathcal{M}}k} \equiv V_{k\mathcal{M}}^{*}$.
Note, that $\hat X_{\bar{\mathcal{M}}}=\hat X_\mathcal{M}^\dagger$.

In our previous publication,\cite{hubop} we considered application
of a method originally formulated in Ref.~\onlinecite{Sandalov}, to 
inelastic transport in molecular junctions.  The main object
of interest in this consideration was many-body (Hubbard) Green function
on the Keldysh contour
\begin{equation}
 \label{GF}
 G_{(a;b),(c;d)}(\tau,\tau') \equiv
 -i <T_c \hat X_{ab}(\tau)\,\hat X_{cd}^\dagger(\tau')>
\end{equation}
where $a, b, c, d$ are many-body states of an isolated molecule,
$T_c$ is the contour ordering operator, and $\tau$, $\tau'$
are contour variables. The consideration
leads to a formulation similar to standard diagrammatic technique,
with series of functional derivatives in place of expansion in small parameter
for conventional diagrammatic consideration (for a detailed discussion
see Ref.~\onlinecite{hubop}). Thus obtained machinery is quite general, but
may be too heavy for realistic applications.  Also at approximate level of 
treatment it may lead to non-physical results.\cite{Novotny,Novotny2}

Here the main object of interest will be the operator 
$\hat X_{ab}(t) = e^{i\hat H t}\hat X_{ab} e^{-i\hat H t}$,
where $|a> \equiv |N_a,s_a>$ and $|b> \equiv |N_b,s_b>$ are 
many-body states defined in (\ref{orthonormal}), and
$t$ is time. Our goal is utilizing Green function techniques find a
(approximate) connection to density matrix based considerations
(in a manner similar to that of Ref.~\onlinecite{Tretiak}), and use resulting
scheme as a simplified version of a procedure considered in e.g. 
Ref.~\onlinecite{hubop}. Note, that we are going to go beyond standard 
QME considerations of transport (see discussion below).
Our starting point is EOM
\begin{equation}
 \label{Heisenberg}
 \frac{<\hat X_{ab}(t)>}{dt} = i\left<\left[\hat H;\hat X_{ab}(t)\right]\right>
\end{equation}
Taking commutator in the right side of (\ref{Heisenberg}) yields
correlation functions of the form (for detailed derivation see
Appendix~\ref{appA}) $<\hat X^\dagger_{(\ldots)}(t)\,\hat c_k(t)>$ and
$<\hat c_k^\dagger(t)\,\hat X_{(\ldots)}(t)>$. 
As usual\cite{MeirWingreen,JauhoWingreenMeir}
these correlation functions can be treated as lesser projections 
of Green functions
\begin{align}
 \label{GcX}
 G_{cX}(\tau,\tau') &= -i<T_c\hat c_k(\tau)\,\hat X_{(\ldots)}^\dagger(\tau')>
 \\
 \label{GXc}
 G_{Xc}(\tau,\tau') &= -i<T_c\hat X_{(\ldots)}(\tau)\,\hat c_k^\dagger(\tau')>
\end{align}
respectively, taken at equal time. The last can be obtained by applying
Langreth projection rules\cite{Langreth} to on-the-contour EOMs for 
(\ref{GcX}) and (\ref{GXc})
\begin{align}
 \label{GcX_EOM}
 G_{cX}(\tau,\tau') &= \sum_\mathcal{M} \int_c d\tau_1\,
 g_k(\tau,\tau_1)\, V_{k\mathcal{M}}\, G_{\mathcal{M},\ldots}(\tau_1,\tau')
 \\
 \label{GXc_EOM}
 G_{Xc}(\tau,\tau') &= \sum_\mathcal{M} \int_c d\tau_1\,
 G_{\ldots,\mathcal{M}}(\tau,\tau_1)\, V_{\bar{\mathcal{M}}k}\, g_k(\tau_1,\tau')
\end{align}
where $G_{\mathcal{M},\ldots}$ ($G_{\ldots,\mathcal{M}}$) is defined in 
(\ref{GF}), $V_{k\mathcal{M}}$ ($V_{\bar{\mathcal{M}}k}$) is introduced in 
(\ref{VME}), and 
\begin{equation}
 \label{gk}
 g_k(\tau,\tau') \equiv -i<T_c c_k(\tau)\, c_k^\dagger(\tau')>
\end{equation}
is Green function of free electrons in the contacts.

Using lesser projections taken at equal times of (\ref{GcX_EOM}) and 
(\ref{GXc_EOM}) in (\ref{Heisenberg}) leads to (see Appendix~\ref{appA}
for details)
\begin{align}
 \label{X_EOM}
 &\frac{d<\hat X_{ab}(t)>}{dt} =
 i\sum_{s}\left[H^{(N_a)}_{ss_a}<\hat X_{(N_a,s;N_b,s_b}(t)>
 \right. \nonumber \\
 &\left. -H_{s_bs}^{(N_b)}<\hat X_{N_a,s_a;N_b,s}(t)>\right]
 + \sum_{\mathcal{M},s} \int_{-\infty}^t dt_1\,
 \nonumber \\
 & \left\{\ G^{<}_{(N_a,s_a;N_b+1,s),\mathcal{M}}(t,t_1)\,  
   \Sigma^{>}_{\mathcal{M},(N_b,s_b;N_b+1,s)}(t_1-t)
 \right.\nonumber \\
 &+ \Sigma^{>}_{(N_a,s_a;N_a+1,s),\mathcal{M}}(t-t_1)\,
    G^{<}_{\mathcal{M},(N_b,s_b;N_a+1,s)}(t_1,t)
 \nonumber \\
 & - G^{>}_{(N_a,s_a;N_b+1,s),\mathcal{M}}(t,t_1)\,
     \Sigma^{<}_{\mathcal{M},(N_b,s_b;N_b+1,s)}(t_1-t)
 \nonumber \\
 & - \Sigma^{<}_{(N_a,s_a;N_a+1,s),\mathcal{M}}(t-t_1)\,
     G^{>}_{\mathcal{M},(N_b,s_b;N_a+1,s)}(t_1,t)
 \nonumber \\
 & - (-1)^{N_a-N_b} \times
 \\
 & \left[\ G^{<}_{(N_a-1,s;N_b,s_b),\mathcal{M}}(t,t_1)\,
     \Sigma^{>}_{\mathcal{M},(N_a-1,s;N_a,s_a)}(t_1-t)
   \right.
 \nonumber \\
 & + \Sigma^{>}_{(N_b-1,s;N_b,s_b),\mathcal{M}}(t-t_1)\,
     G^{<}_{\mathcal{M},(N_b,s_b;N_a+1,s)}(t_1,t)
 \nonumber \\
 & - G^{>}_{(N_a-1,s;N_b,s_b),\mathcal{M}}(t,t_1)\,
     \Sigma^{<}_{\mathcal{M},(N_a-1,s;N_a,s_a)}(t_1-t)
 \nonumber \\
 & \left.\left.
   - \Sigma^{<}_{(N_b-1,s;N_b,s_b),\mathcal{M}}(t-t_1)\,
     G^{>}_{\mathcal{M},(N_b-1,s;N_a,s_a)}(t_1,t)
   \right]\right\}
 \nonumber 
\end{align}
Here $\Sigma^{>,<}_{\mathcal{M}_1,\mathcal{M}_2}(t)$
are greater and lesser molecular self-energies due to coupling to the contacts
\begin{align}
 \label{SE_gtlt}
 \Sigma^{>,<}_{\mathcal{M}_1,\mathcal{M}_2}(t) =&
 \sum_{K=L,R}\Sigma^{(K)>,<}_{\mathcal{M}_1,\mathcal{M}_2}(t)
 \\
 \label{SEK_gtlt}
 \Sigma^{(K)>,<}_{\mathcal{M}_1,\mathcal{M}_2}(t) =& \sum_{k\in K}
 V_{\bar{\mathcal{M}}_1,k}\, g_k^{>,<}(t)\, V_{k,\mathcal{M}_2}
\end{align}
with $g_k^{>,<}(t)$ being greater and lesser projections of (\ref{gk})
\begin{align}
 \label{gk_gt}
 g_k^{>}(t) =& -i [1-n_k] e^{-i\varepsilon_k t}
 \\
 \label{gk_lt}
 g_k^{<}(t) =& i n_k e^{-i\varepsilon_k t}
\end{align}
and $G^{>,<}_{\mathcal{M}_1,\mathcal{M}_2}(t_1,t_2)$ are
greater and lesser projections of (\ref{GF})
\begin{align}
 \label{Ggt}
 G^{>}_{(a;b),(c;d)}(t_1,t_2) =&
 -i < \hat X_{ab}(t_1)\, \hat X_{cd}^\dagger(t_2) >
 \\
 \label{Glt}
 G^{<}_{(a;b),(c;d)}(t_1,t_2) =&
 \pm i < \hat X_{cd}^\dagger(t_2)\,\hat X_{ab}(t_1) >
\end{align}
Note, in (\ref{Glt}) `${}+{}$' stands when both $\mathcal{M}_1$
and $\mathcal{M}_2$ are transitions of Fermi type, and `${}-{}$' otherwise.
For future reference we also define a damping matrix in Liouville space
\begin{equation}
 \label{Gamma}
 \Gamma^{(K)}_{\mathcal{M}_1,\mathcal{M}_2} \equiv
 i\left[\Sigma^{>}_{\mathcal{M}_1,\mathcal{M}_2}
       -\Sigma^{<}_{\mathcal{M}_1,\mathcal{M}_2}\right]
\end{equation}

Expression for the current can be derived in a similar way
(see Eq.(10) of Ref.~\onlinecite{hubop})
\begin{align}
 \label{IK}
 I_K(t) &= \frac{e}{\hbar} \sum_{\mathcal{M},\mathcal{M}'}
 \int_{-\infty}^{t}dt'
 \nonumber \\
 & \left\{\ \Sigma^{<}_{\mathcal{M},\mathcal{M}'}(t-t')\,
   G^{>}_{\mathcal{M}',\mathcal{M}}(t',t)
 \right. \nonumber \\
 &+ G^{>}_{\mathcal{M},\mathcal{M}'}(t,t')\,
   \Sigma^{<}_{\mathcal{M}',\mathcal{M}}(t'-t)
 \\
 & 
 - \Sigma^{>}_{\mathcal{M},\mathcal{M}'}(t-t')\,
   G^{<}_{\mathcal{M}',\mathcal{M}}(t',t)
 \nonumber \\ & \left.
 - G^{<}_{\mathcal{M},\mathcal{M}'}(t,t')\,
   \Sigma^{>}_{\mathcal{M}',\mathcal{M}}(t'-t)
 \right\}
 \nonumber
\end{align}

Equations (\ref{X_EOM}) and (\ref{IK}) are exact, however their
right sides are expressed in terms of Green functions. Our goal now
is to introduce approximate scheme in order to close (\ref{X_EOM})
in terms of $<\hat X_{ab}(t)>$, thus finding connection to QME.
This approximation is introduced and discussed in the next Section.

\section{\label{qme}Generalized QME}

Before introduction of an ansatz to close Eq.(\ref{X_EOM}) we note
close connection between EOM for $\hat X_{ab}(t)$ and density matrix
element $\rho_{ba}(t)$. Indeed, 
\begin{align}
 \rho_{ba}(t) &= \ll \hat X_{ba}|e^{-i\mathcal{L}t}|\hat \rho_0\gg 
               = \ll \hat X_{ba}|e^{-i\mathcal{L}t}\hat\rho_0\gg
 \\
              &= \ll e^{i\mathcal{L}^\dagger t}\hat X_{ba}|\hat\rho_0\gg 
               = <\hat X_{ab}(t)> 
 \nonumber
\end{align}
where $\mathcal{L}$ is the total Liouvillian and 
$\ll A|B\gg \equiv\mbox{Tr}[\hat A^\dagger\,\hat B]$
is scalar product in Liouville space. Hence we expect that Eq.(\ref{X_EOM})
after introducing approximation expressing it in terms of $\hat X_{ab}$ 
only should result in QME.

Correlation function of the type (\ref{Ggt}) can be 
{\em exactly} written in Liouville space as  
\begin{align}
\label{GKBAusingAns1}
 &<\hat X_{ab}(t_1)\,\hat X_{cd}^\dagger(t_2)> =
 \\ 
 & \quad \theta(t_1-t_2) 
   \ll \hat X_{ba} \hat{I}_{K} | \mbox{e}^{-i\mathcal{L}(t_1-t_2)} | \hat X_{dc} \hat \rho(t_2) \gg
 \nonumber \\
 & + \theta(t_2-t_1)
   \ll \hat X_{cd} \hat{I}_{K} | \mbox{e}^{-i\mathcal{L}(t_2-t_1)} | \hat \rho(t_1) \hat X_{ab}  \gg.
 \nonumber
\end{align}  
We introduce the projector superoperator 
\begin{equation}
\label{Projector}
{\cal P} = \sum_{ef} | \hat X_{ef} \hat \rho_{K}^{eq} \gg \ll \hat X_{ef} \hat I_{K} |,
\end{equation}  
which disregards nonequilibrium features in the leads and decouples system and
bath dynamics.
The ansatz, we propose, replaces (\ref{GKBAusingAns1}) by
\begin{align}
\label{GKBAusingAns2}
 &<\hat X_{ab}(t_1)\,\hat X_{cd}^\dagger(t_2)> \approx
 \\ 
 & \quad \theta(t_1-t_2) 
   \ll \hat X_{ba} \hat{I}_{K} | \mbox{e}^{-i\mathcal{L}(t_1-t_2)} {\cal P}  | \hat X_{dc} \hat \rho(t_2) \gg
 \nonumber \\
 & + \theta(t_2-t_1)
   \ll \hat X_{cd} \hat{I}_{K} | \mbox{e}^{-i\mathcal{L}(t_2-t_1)} {\cal P}  | \hat \rho(t_1) \hat X_{ab}  \gg.
 \nonumber
\end{align}  
Next we introduce retarded and advanced Green functions in the Liouville space
(see Appendix~\ref{appB})
\begin{align}
 \label{GrNew}
 \mathcal{G}^r_{ij,mn}(t) &\equiv -i\theta(t) 
 \ll \hat X_{ji} \hat{I}_{K} | e^{-i \mathcal{L} t} | \hat X_{nm} \hat \rho_{K}^{eq} \gg
 \\ \nonumber
 & = -i\theta(t) \ll \hat X_{ji}| \mathcal{U}_{eff}(t) | \hat X_{nm} \gg \\
 \label{GaNew}
 \mathcal{G}^a_{ij,mn}(t) &\equiv i \theta(-t)
 \ll \hat X_{mn} \hat{I}_{K} | e^{i \mathcal{L} t} | \hat X_{ij} \hat \rho_{K}^{eq}  \gg 
 \\ \nonumber
 & = i \theta(-t) \ll \hat X_{ji} | \mathcal{U}_{eff}^{\dagger}(-t) | \hat X_{nm} \gg  \nonumber
\end{align}
where the effective propagator in the molecule space reads
\begin{equation}
\label{EffPropagator}
\mathcal{U}_{eff} \equiv
\ll \cdot \hat{I}_{K} | \mbox{e}^{-i\mathcal{L}t} | \cdot \hat \rho_{K}^{eq} \gg 
= \mbox{Tr}_{K} \{ e^{-i \mathcal{L} t} \hat \rho_{K}^{eq} \}.
\end{equation}
Using (\ref{GrNew}) and (\ref{GaNew}), we can rewrite (\ref{GKBAusingAns2}) as
\begin{align}
 \label{GKBA}
 &<\hat X_{ab}(t_1)\,\hat X_{cd}^\dagger(t_2)> = 
 \nonumber \\
 & i\sum_{e,f}\left[ 
    \mathcal{G}^r_{ab,fe}(t_1-t_2) <\hat X_{fe}(t_2)\hat X_{cd}^\dagger(t_2)>
 \right. \nonumber \\ & \left. \qquad
    -<\hat X_{ab}(t_1)\hat X_{ef}^\dagger(t_1)>\mathcal{G}^a_{ef,cd}(t_1-t_2)
    \right]
 \\
 & \equiv i \sum_m \left[
   \mathcal{G}^r_{ab,md}(t_1-t_2) <\hat X_{mc}(t_2)>
 \right. \nonumber \\ &\left. \qquad
  -<\hat X_{am}(t_1)>\mathcal{G}^a_{mb,cd}(t_1-t_2)
 \right]
 \nonumber
\end{align}  
where second equality follows from orthonormality condition 
(\ref{orthonormal}).

Similar consideration for correlation function (\ref{Glt}) leads to
\begin{align}
 \label{GKBA2}
 &<\hat X_{cd}^\dagger(t_2)\,\hat X_{ab}(t_1)> = 
 \nonumber \\
 & i\sum_{e,f}\left[ 
    \mathcal{G}^r_{ab,fe}(t_1-t_2) <\hat X_{cd}^\dagger(t_2)\hat X_{fe}(t_2)>
 \right. \nonumber \\ &\left. \qquad
    -<\hat X_{ef}^\dagger(t_1)\hat X_{ab}(t_1)>\mathcal{G}^a_{ef,cd}(t_1-t_2)
    \right]
 \\
 & \equiv i \sum_m \left[
   \mathcal{G}^r_{ab,cm}(t_1-t_2) <\hat X_{dm}(t_2)>
 \right. \nonumber \\ &\left. \qquad
  -<\hat X_{mb}(t_1)>\mathcal{G}^a_{am,cd}(t_1-t_2)
 \right]
 \nonumber 
\end{align}  
It is interesting to note that (\ref{GKBA}) and (\ref{GKBA2}) can be 
considered as the Liouville space analog of the generalized Kadanoff-Baym 
ansatz.\cite{HaugJauho}

Using (\ref{GKBA}) and (\ref{GKBA2}) in (\ref{X_EOM}) closes the latter 
equation in terms of DM $\rho_{ba}(t)\equiv <\hat X_{ab}(t)>$ only
\begin{widetext}
\begin{align}
\label{gen_QME}
 \frac{d\rho_{12}(t)}{dt} =& -i\sum_{3,4}\left\{
 \delta_{N_1,N_3}\delta_{N2,N_4}
 \sum_s\left(\delta_{i_2,i_4}H_{i_1,i_3}^{(N_1)}
            -\delta_{i_1,i_3}H_{i_4,i_2}^{(N_2)}\right)
 \right.\nonumber \\
 & -i\sum_{s_1,s_2}\int_{-\infty}^{+\infty}dt_1
\left[ \mathcal{G}^r_{(2;N_1+1,s_1)(4;N_3+1,s_2)}(t-t_1)
         \Sigma^{<}_{(3;N_3+1,s_2)(1;N_1+1,s_1)}(t_1-t)
  \right.
 \nonumber \\ &\qquad\qquad\qquad\quad
        -\Sigma^{<}_{(2;N_2+1,s_1)(4;N_4+1,s_2)}(t-t_1)
         \mathcal{G}^a_{(3;N_4+1,s_2)(1;N_2+1,s_1)}(t_1-1)
 \nonumber \\ &\qquad\qquad\qquad\quad
        -\mathcal{G}^r_{(N_2-1,s_1;1)(N_4-1,s_2;3)}(t-t_1)
         \Sigma^{>}_{(N_4-1,s_2;4)(N_2-1,s_1;2)}(t_1-t)
 \nonumber \\ &\qquad\qquad\qquad\quad
        +\Sigma^{>}_{(N_1-1,s_1;1)(N_3-1,s_2;3)}(t-t_1)
         \mathcal{G}^a_{(N_3-1,s_2;4)(N_1-1,s_1,2)}(t_1-t)
 \nonumber \\ &\ \ 
 -(-1)^{N_1-N_2}\times
  \left(\mathcal{G}^r_{(N_2-1,s_1;1)(4;N_3+1,s_2)}(t-t_1)
        \Sigma^{<}_{(3;N_3+1,s_2)(N_2-1,s_1;2)}(t_1-t)
  \right.
  \\ &\qquad\qquad\qquad\quad
       -\Sigma^{<}_{(N_1-1,s_1;1)(4;N_4+1,s_2)}(t-t_1)
        \mathcal{G}^a_{(3;N_4+1,s_2)(N_1-1,s_1;2)}(t_1-1)
 \nonumber \\ &\qquad\qquad\qquad\quad
       -\mathcal{G}^r_{(2;N_1+1,s_1)(N_4-1,s_2;3)}(t-t_1)
        \Sigma^{>}_{(N_4-1,s_2;4)(1;N_1+1,s_1)}(t_1-t)
 \nonumber \\ &\qquad\qquad\qquad\quad
   \left. \left. \left.
       +\Sigma^{>}_{(2;N_2+1,s_1)(N_3-1,s_2;3)}(t-t_1)
        \mathcal{G}^a_{(N_3-1,s_2;4)(1;N_s+1,s_1)}(t_1-t)
   \right)\right]
   \times \rho_{34}(t_1) \right\}
 \nonumber
\end{align}
\end{widetext}
This is a generalized non-Markovian QME. Note, that prefactor $(-1)^{N_1-N-2}$
coming from coherences between different charge blocks is usually lost in 
the standard QME derivations.

To make (\ref{gen_QME}) more tractable below we assume
Markovian generator $\mathcal{L}_{eff}$ 
(e.g. Markovian Redfield generator\cite{Breuer02,Gardiner00,KuboB98b})
for retarded and advanced Green functions 
(\ref{GrNew}) and (\ref{GaNew})
\begin{equation}
\label{MarkPropa}
\mathcal{U}_{eff}(t) \approx e^{-i \mathcal{L}_{eff} t},
\end{equation}
So that
\begin{align}
 \label{Gr}
 \mathcal{G}^r_{ij,mn}(t) &\equiv -i\theta(t) 
 \ll \hat X_{ji} | e^{-i \mathcal{L}_{eff} t} | \hat X_{nm} \gg
 \\
 \label{Ga}
 \mathcal{G}^a_{ij,mn}(t) &\equiv i \theta(-t)
 \ll \hat X_{ji} | e^{-i \mathcal{L}_{eff}^\dagger t} | \hat X_{nm} \gg.
\end{align}
The ansatz (\ref{GKBAusingAns2}) together with (\ref{Gr}) and (\ref{Ga}) is 
equivalent to use of the regression formula on the Hubbard Green function. 
This procedure is commonly used to calculate multipoint correlation 
functions using effective Markovian 
propagators\cite{Breuer02,Gardiner00,Carmichael}.

The standard non-Markovian QME\cite{HarbolaEsposito06} is obtained from 
(\ref{gen_QME}) by using in (\ref{Gr}) and (\ref{Ga}) 
the free molecular evolution 
$\mathcal{L_M} = [\hat H_M,\cdot]$ instead of the effective one 
$\mathcal{L}_{eff}$. 
Note that difference between standard and generalized versions of QME
is similar to dressing of diagrams in GF diagrammatic technique. 
Note also that the standard QME by itself can not reproduce, e.g., broadening 
of molecular levels due to coupling to the contacts as noted in 
Ref.~\onlinecite{Wacker05}.

Below we use the Markovian Redfield equation to get $\mathcal{L}_{eff}$ 
(see Appendix~\ref{appC}).  Its spectral decomposition
\begin{equation}
 \label{spectral}
 \mathcal{L}_{eff} = \sum_\gamma |R_\gamma\gg  \lambda_\gamma \ll L_\gamma|
\end{equation}
with eigenvalues $\lambda_\gamma$ and left $|L_\gamma\gg $ and right $|R_\gamma\gg $ eigenvectors,
provides a numerically tractable scheme to deal with generalized QME (\ref{gen_QME}) by utilizing
\begin{align}
 \mathcal{G}^r_{ij,mn}(t) &= -i\theta(t) \sum_\gamma 
 \ll ji | R_\gamma \gg e^{-i\lambda_\gamma t} \ll L_\gamma | nm\gg
 \\
 \mathcal{G}^a_{ij,mn}(t) &= i\theta(-t) \sum_\gamma
 \ll ji | L_\gamma\gg e^{-i\lambda_\gamma^{*}t}\ll R_\gamma|nm\gg
\end{align}
Steady-state for (\ref{gen_QME}) is given by the right eigenvector
with zero eigenvalue of the Liouvillian corresponding to the Markov limit
of (\ref{gen_QME}).

Similarly, approximate expression for current in terms of $<\hat X_{(\ldots)}>$ 
can be obtained using (\ref{GKBA}) and (\ref{GKBA2}) in (\ref{IK})
\begin{align}
 \label{IK_GKBA}
 I_K(t) &= \frac{e}{\hbar} \sum_{\mathcal{M}_1,\mathcal{M}_2} 
 \sum_e \int_{-\infty}^{+\infty} dt_1
 \nonumber\\
  2\mbox{Re} & \left[
  \mathcal{G}^r_{(N_1,i_1;N_1+1,j_1),(N_2,i_2;e)}(t-t_1)\,
 \right. \nonumber \\
 & \times
  \Sigma^{>}_{(N_2,i_2;N_2+1,j_2),(N_1,i_1;N_1+1,j_1)}(t_1-t)
 \nonumber \\
 & \times
  <\hat X_{(N_2+1,j_2;e)}(t_1)>
 \\
 & +
  \mathcal{G}^r_{(N_1,i_1;N_1+1,j_1),(e;N_2+1,j_2)}(t-t_1)\,
 \nonumber \\
 & \times
  \Sigma^{<}_{(N_2,i_2;N_2+1,j_2),(N_1,i_1;N_1+1,j_1)}(t_1-t)
 \nonumber \\
 & \left. \times
  <\hat X_{(e;N_2,i_2)}(t_1)> 
 \right]
 \nonumber
\end{align}

\section{\label{results}Results and discussion}
As a first example we consider a simple resonant level model. 
One has two charge blocks (occupied and unoccupied level) 
with one state in each of them: $|0>$ and $|1>$.
The molecular Hamiltonian is $\hat H_M=|1>\varepsilon_0<1|$.
Current (\ref{IK_GKBA}) in this case becomes
\begin{align}
 \label{IK_rlm}
 &I_K(t) = \frac{ie}{\hbar} \int_{-\infty}^t dt_1\,\left\{ 
 \right. \\
 & \quad \left[ \mathcal{G}^r_{01,01}(t-t_1)\Sigma_K^{>}(t_1-t)
       - \Sigma_K^{>}(t-t_1)\mathcal{G}^a_{01,01}(t_1-t)
  \right] 
 \nonumber \\ &\quad\times\rho_{11}(t_1)
 \nonumber \\ &
 +\left[ \mathcal{G}^r_{01,01}(t-t_1)\Sigma_K^{<}(t_1-t)
       - \Sigma_K^{<}(t-t_1)\mathcal{G}^a_{01,01}(t_1-t)
  \right] 
 \nonumber \\ &\left.\quad\times\rho_{00}(t_1)
 \right\}
 \nonumber
\end{align}
where 
\begin{align}
 \label{Gr0110}
 \mathcal{G}^r_{01,01}(t) &= -i\theta(t)e^{-i(\varepsilon_0-i\Gamma/2)t}
 \equiv G^r(t)
 \\
 \label{Ga1001}
 \mathcal{G}^a_{01,01}(t) &= i\theta(-t)e^{-i(\varepsilon_0+i\Gamma/2)t}
 \equiv G^a(t)
\end{align}
and $\Gamma=\sum_{K=L,R}\Gamma^K_{01,01}$ with $\Gamma^K$ defined in 
(\ref{Gamma}).\\
Generalized QME (\ref{X_EOM}) yields
\begin{align}
 \label{rho_rlm}
 &\frac{d\rho_{11}(t)}{dt} = -\frac{d\rho_{00}(t)}{dt} 
 = \int_{-\infty}^{t}dt_1\, \left\{
 \right. \\
 &\quad \left[ \mathcal{G}^r_{01,01}(t-t_1)\Sigma^{>}(t_1-t)
       - \Sigma^{>}(t-t_1)\mathcal{G}^a_{01,01}(t_1-t)
  \right] 
 \nonumber \\ &\quad\times\rho_{11}(t_1)
 \nonumber \\ &
 +\left[ \mathcal{G}^r_{01,01}(t-t_1)\Sigma^{<}(t_1-t)
       - \Sigma^{<}(t-t_1)\mathcal{G}^a_{01,01}(t_1-t)
  \right] 
 \nonumber \\ &\left.\quad\times
 \rho_{00}(t_1)
 \right\}
 \nonumber
\end{align}
At steady-state (\ref{rho_rlm}) yields
\begin{equation}
 \label{rho_rlm_ss}
 \rho_{11} = 1-\rho_{00} = n_0
\end{equation}
with $n_0$ average occupation of the level
\begin{align}
 n_0 &= \int_{-\infty}^{+\infty}\frac{dE}{2\pi} A(E)
 \left[\frac{\Gamma_L}{\Gamma}f_L(E)+\frac{\Gamma_R}{\Gamma}f_R(E)\right]
 \\
 A(E) &= \frac{\Gamma}{(E-\varepsilon_0)^2+(\Gamma/2)^2}
\end{align}
where $A(E)$ is spectral function and $f_K(E)$ is Fermi distribution in
contact $K=L,R$. Using (\ref{Gr0110}), (\ref{Ga1001}), 
and (\ref{rho_rlm_ss}) in (\ref{IK_rlm}) leads to the Landauer expression
\begin{equation}
 \label{Landauer}
 I_K = \frac{e}{\hbar}\int_{-\infty}^{+\infty}\frac{dE}{2\pi}\,
 \frac{\Gamma_L\Gamma_R}{\Gamma}A(E)[f_L(E)-f_R(E)] 
\end{equation}
Note, that generalized QME approach takes level broadening into account
in a natural way contrary to the standard QME considerations. 

\begin{figure}[htbp]
\centering\includegraphics[width=\linewidth]{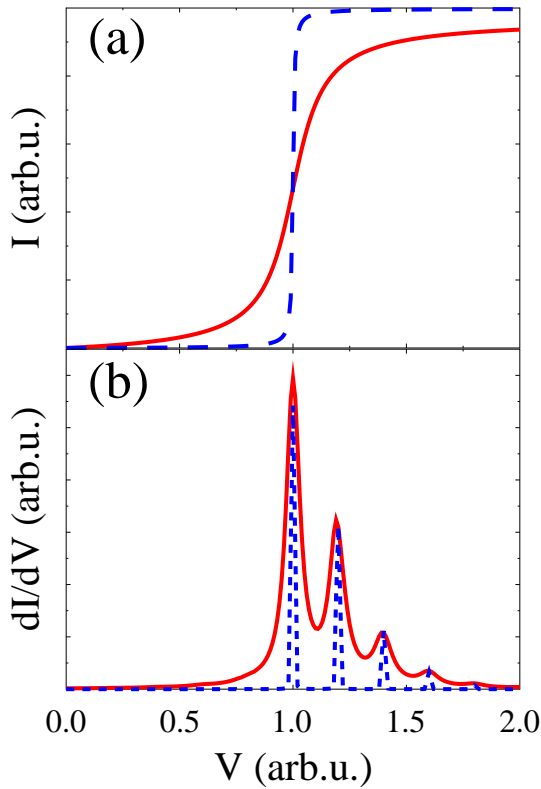}
\caption{\label{f1}
(Color online)
Comparison of generalized (solid line, red) to standard (dashed line, blue)
QME  results for resonant level model. 
(a) Current vs. bias for single level.
(b) Conductance vs. bias for single level coupled to a vibration.
See text for parameters.
}
\end{figure}

Now we present several numerical examples.
Figure~\ref{f1} compares results of calculation within
our generalized QME (solid line) and standard QME (dashed line)
approaches. Fig.~\ref{f1}a shows current-voltage characteristic
of single resonant level $\varepsilon_0$ model. Generalized QME accounts for
level broadening due to coupling to the contacts, while standard QME
approach misses the broadening altogether. Parameters of the calculation 
are $\varepsilon_0=1$, $\Gamma_L=\Gamma_R=0.1$, $E_F=\mu_R=0$, 
$\mu_L=E_F+|e|V_{sd}$. Here and below we use arbitrary units.
Fig.~\ref{f1}b shows conductance vs. bias for the model of single
level $\varepsilon_0$ coupled to a vibration $\omega_0$. Once more,
while generalized QME provides reasonable results (compare e.g to Fig.4
of Ref.~\onlinecite{strong_elph})a standard QME approach is 
capable of prediciting only positions of the peaks. 
Parameters of the calculation are $\Gamma_L=\Gamma_R=0.05$,
$\omega_0=0.2$, and $M=0.2$. The last is strenth of electron-vibration 
coupling on the bridge with corresponding Hamiltonian  
$M(\hat a+\hat a^\dagger)\hat n_0$, where $\hat a^\dagger$ ($\hat a$)
are creation (annihilation) operators of vibrational quanta and
$\hat n_0$ operator of the evel population. Other parameters are as in 
Fig.~\ref{f1}a. 
Note, that in simulations we used small but finite broadening for the standard
QME approach in order to avoid delta-function divergencies in conductunce.
We also scaled the standard QME result in Fig.~\ref{f1}b for convenience.

\begin{figure}[htbp]
\centering\includegraphics[width=\linewidth]{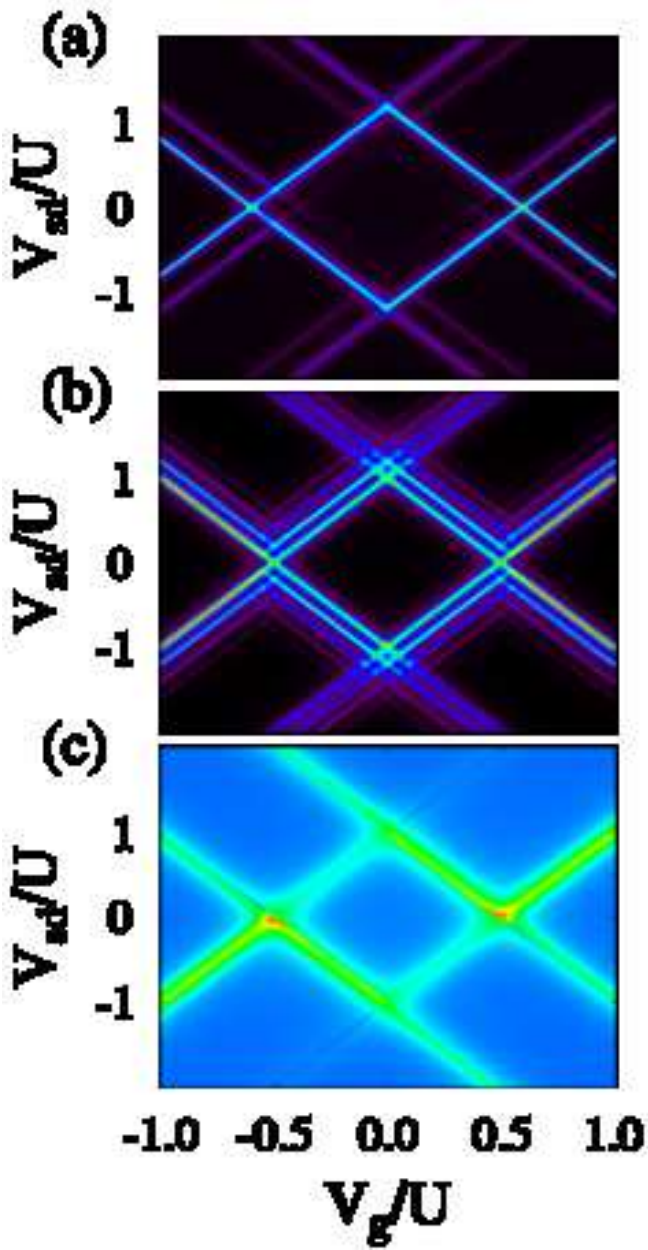}
\caption{\label{f2}
(Color online)
Conductance vs. applied bias $V_{sd}$ and gate voltage $V_g$
for a quantum dot (QD) within generalized QME approach. 
Shown are results for models of (a) QD with level degenraucy removed
(e.g. by applied magnetic field), (b) QD coupled to a vibration,
and (c) QD with asymmetric coupling to the contacts.
See text for parameters. 
}
\end{figure}

Figures~\ref{f2} present conductance maps for a quantum dot 
obtained within the generalized QME approach and similar
to those obtained within many-body Green function technique
(see Ref.~\onlinecite{hubop}). Parameters of the calculation
are level positions $\varepsilon_\sigma=-0.5$ 
($\sigma=\{\uparrow,\downarrow\}$),
molecule-contacts coupling $\Gamma_{K,\sigma}=0.01$ ($K=\{L,R\}$),
on-site repulsion $U=1$, Fermi level $E_F=0$. Electrochemical potentials
in the contacts are $\mu_L=E_F+|e|V_{sd}/2$ and $\mu_R=E_F-|e|V_{sd}/2$.
Deviations from this `standard' set for each calculation are specified below.
Fig.~\ref{f2}a shows conductance map for QD with level degeneracy removed by
e.g. external magnetic field $\varepsilon_\uparrow=-0.6$ and 
$\varepsilon_\downarrow=-0.4$. For discussion on origin and intensity of 
peaks see Ref.~\onlinecite{CB}. Fig.~\ref{f2}b shows conductance
map for QD coupled to a vibration $\omega_0=0.1$ and $M=0.1$.
In addition to elastic peaks vibrational sidebands corresponding to resonant
inelastic tunneling are reproduced as well.
Fig.~\ref{f2}c shows conductance map for QD with asymmetric 
coupling to the contacts $\Gamma_{L,\sigma}=0.01$ and
$\Gamma_{R,\sigma}=0.1$. This result is similar to the one presented
in Fig.4 of Ref.~\onlinecite{Wacker05}

Note, that vibrations in both Fig.~\ref{f1}b and \ref{f2}b were
introduced, as is usually done in resonant inelastic transport conѕiderations,
with the help of small polaron transformation. So that vibrational features
in electron transport stem from the Franck-Condon factors calculated 
under assumption of unperturbed thermal distribution of vibrational
population. Actual vibrational states are not included in the
current consideration, and their incorporation into many-body state
description will be described elsewhere. 

\begin{figure}[htbp]
\centering\includegraphics[width=\linewidth]{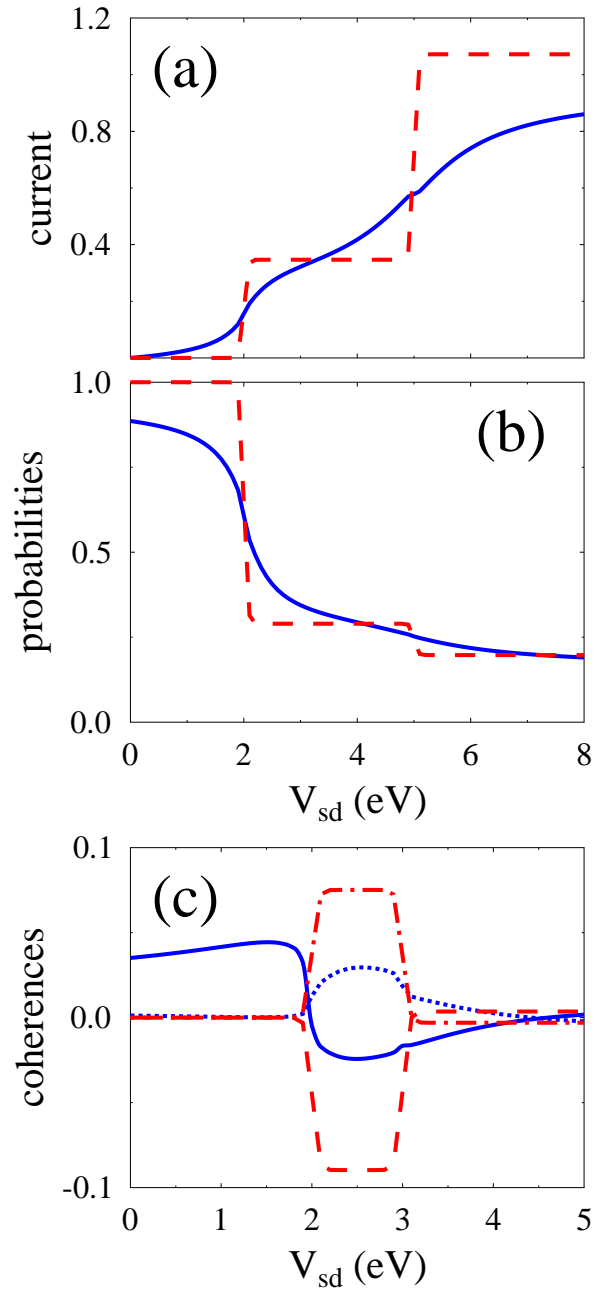}
\caption{\label{f3}
(Color online)
Two-level bridge with coherences in the eigenbasis of the system 
induced by coupling to contacts.\cite{HarbolaEsposito06} 
(a) Current and 
(b) probability for the system to be unoccupied
vs. aplied bias -- generalized (solid line, blue)
and standard (dashed line, red) QME considerations.
(c) Real (solid, blue and dashed, red) and 
imaginary (dotted, blue and dash-dotted, red) parts of
coherences in the local basis vs. applied bias
for generalized and standard QME
treatment, respectively. 
See text for parameters.
}
\end{figure}

Finally, we consider a model of two-level bridge with coherences
in the eigenbasis of the bridge induced by coupling to the contacts.
This models was previously considered in Ref.~\onlinecite{HarbolaEsposito06}
within standard QME approach. Figure~\ref{f3} presents comparison
between standard and generalized QME approaches. Parameters of the
calculation are similar to those in Ref.~\cite{HarbolaEsposito06} --
eigenenergies of the bridge are $\varepsilon_1=5$eV are 
$\varepsilon_2$eV, strength of their coupling to contacts is 
$T_1^L=T_2^L=0.3$eV, $T_1^R=0.2$eV, $T_2^R=0.4$eV.
For temperature we take physically reasonable value of $T=0.03$eV.
Figures~\ref{f3}a and \ref{f3}b show current and one of the probabilities
(probability of the system to be unoccupied) vs. applied bias. 
One sees that broadening due to coupling to the contacts is
preserved in our scheme. Note, that broadening presented in 
Ref.~\onlinecite{HarbolaEsposito06} was due to unphysically high value of 
temperature chosen. Fig.~\ref{f3}c demonstrateѕ influence of broadening
on coherences (in local basis). Here we bring the two eigenenergies 
closer to each other, $\varepsilon_1=3$eV, in order to make coherences due
to coupling to the contacts more pronounced. One sees that taking level 
broadening into account changes the coherences essentially.

\section{\label{conclude}Conclusion}
Necessity for description of molecular transport in the language of 
many-body (isolated molecule) states, essential for description
of resonant tunneling and for study of optoelectronic devices,
has been realized and several approaches were 
proposed.\cite{Datta,Datta_ph,Koch2005,Koch2006,Koch2006a,Koch2007,Petrov2005,Petrov2006,Sandalov,Fransson,hubop}
Here we introduce a simplified version of the Hubbard operator Green
function approach considered in application to inelastic transport
in our previous publication.\cite{hubop}
The simplified approach is formulated for density matrix instead of GF 
and provides more easy way for calculating both time-dependent and
steady state transport in molecular junctions. 
Starting from GF-type consideration we introduce Liouville
space analog of the generalized Kadanoff-Baym ansatz, which allows us 
to derive generalized QME. The latter differs from the standard QME
by incorporating effective propagation in place of free evolution.
The procedure is similar in spirit to diagrams dressing in 
GF diagrammatic techniques. Capabilities of the scheme are demonstrated 
within model calculations. Application of the approach to opto-electronic
response of molecular junctions is a goal for future research.

\begin{acknowledgments}
M.E. is funded by the FNRS Belgium (charg\'e de recherche) and by
the Luxembourgish Government (bourse de formation-recherche).
M.G. gratefully acknowledges support from the UCSD Startup
Fund. This work was performed, in part, at the Center for Integrated 
Nanotechnologies, a U.S. Department of Energy, Office of Basic Energy 
Sciences user facility at Los Alamos National Laboratory 
(Contract DE-AC52-06NA25396).
\end{acknowledgments}

\vspace*{0.5cm}
\appendix
\section{\label{appA}Derivation of Eq.(\ref{X_EOM})}
We start from Eq.(\ref{Heisenberg}), which after evaluating the commutator
becomes
\begin{align}
 \label{commutator}
 &\frac{d<\hat X_{ab}(t)>}{dt} =
 -i\left\{
   \sum_s\left[ H^{(N_b)}_{s_b,s}<\hat X_{(N_a,s_a;N_b,s)}(t)>
\right.\right. \nonumber \\ & \quad \left.
              - <\hat X_{(N_a,s;N_b,s_b)}>H^{(N_a)}_{s,s_a} \right]
 +\sum_{s,k}\left[ (-1)^{N_a-N_b}\times \right.
 \nonumber \\
 & \left( V_{k,(N_b,s_b;N_b+1,s)} 
   <\hat c_k^\dagger(t)\,\hat X_{(N_a,s_a;N_b+1,s)}(t)>
 \right. \nonumber \\
 & \left. - V_{(N_a+1,s;N_a,s_a),k}
   <\hat X_{(N_b,s_b;N_a+1,s)}^\dagger(t)\,\hat c_k(t)> \right)
 \\
 & + V_{(N_b,s_b;N_b-1,s),k}
   <\hat X_{(N_b-1,s;N_a,s_a)}^\dagger(t)\,\hat c_k(t)>
 \nonumber \\
 & \left.\left. - V_{k,(N_a-1,s;N_a,s_a)}
   <\hat c_k^\dagger(t)\,\hat X_{(N_a-1,s;N_b,s_b)}(t)>
 \right]\right\}
 \nonumber
\end{align}
where $\sum_s\ldots$ is sum over molecular states within charge block,
$\sum_k\ldots$ is sum over states in the contacts, and
factor $(-1)^{N_a-N_b}$ results from commuting $\hat X_{ab}$
with $\hat c_k$ ($\hat c_k^\dagger$).

Correlation functions in the right of Eq.(\ref{commutator})
can be identified as lesser projections of the GFs (\ref{GcX}) and (\ref{GXc})
defined on the Keldysh contour. EOMs for this GFs are presented in 
(\ref{GcX_EOM}) and (\ref{GXc_EOM}). Taking lesser projection of the EOMs
and applying the Langreth rules\cite{Langreth} yields, 
e.g. for the first correlation function in (\ref{commutator})
\begin{align}
 \label{example}
 &<\hat c_k^\dagger(t)\,\hat X_{(N_a,s_a;N_b+1,s)}(t)>
 \equiv (-1)^{N_a-N_b-1}iG_{Xc}(t,t)
 \nonumber \\
 &= (-1)^{N_a-N_b-1}\sum_{\mathcal{M}}\int_{-\infty}^{+\infty}dt_1\,
    \left[(-1)^{N_a-N_b-1} \times
 \right. \nonumber \\
 & \quad <\hat X_{\mathcal{M}}^\dagger(t_1)\,
          \hat X_{(N_a,s_a;N_b+1,s)}(t)> g_k^a(t_1-t)
 \\
 &     +\theta(t-t_1)\left(<\hat X_{(N_a,s_a;N_b+1,s)}(t)\,\,
                            \hat X_{\mathcal{M}}^\dagger(t_1)>
       \right. \nonumber \\ & \quad \left.
          -(-1)^{N_a-N_b-1}<\hat X_{\mathcal{M}}^\dagger(t_1)\,
                            \hat X_{(N_a,s_a;N_b+1,s)}(t)>\right)
  \nonumber \\ & \quad \times \left.
  g_k^{<}(t_1-t)
  \right]
 \nonumber
\end{align}
where $g_k^{a,<}(t)$ are advanced and lesser projections of the GF (\ref{gk}),
$<\ldots>=\mbox{Tr}[\ldots\hat\rho_0]$ with initial density matrix 
taken as usual at infinite past, and where general property of GFs
$G^r(t)=\theta(t)[G^{>}(t)-G^{<}(t)]$ was used for the $G_{XX}$ GF.
Once more factors $(-1)^{N_a-N_b-1}$ trace Fermi or Bose character of 
$\hat X_{ab}$. Using $g_k^a(t)=\theta(-t)[g_k^{<}(t)-g_k^{>}(t)]$
and utilizing (\ref{SE_gtlt}) and (\ref{SEK_gtlt}) 
leads to final expression for the first correlation function in 
(\ref{commutator}). Repeating consideration for the three other
correlation functions in (\ref{commutator}), and using the resulting
expressions in (\ref{commutator}) leads to Eq.(\ref{X_EOM}).

\section{\label{appB}Green functions in the Liouville space}
Here we discuss properties of retarded and advanced Green functions
in the Liouville space. We start from definitions (\ref{GrNew})
and (\ref{EffPropagator}).
Utilizing the property of the full unitary propagator
\begin{equation}
\ll \hat A | \mbox{e}^{-i\mathcal{L}t} | \hat B \gg = 
\ll \hat A^{\dagger} | \mbox{e}^{-i\mathcal{L}t} | \hat B^{\dagger} \gg^*
\end{equation}
one can write
\begin{align}
\label{equality}
\ll \hat X_{ij} | \mathcal{U}_{eff}(t) | \hat X_{mn} \gg &=  
\ll \hat X_{ji} | \mathcal{U}_{eff}(t) | \hat X_{nm} \gg^* \nonumber\\
&= \ll \hat X_{nm} | \mathcal{U}_{eff}^{\dagger}(t) | \hat X_{ji} \gg
\end{align}  
where the second equality comes from definition of Hermitian conjugate.
Using (\ref{equality}) in (\ref{GrNew}) one gets
\begin{align}
 \label{GaGrNew}
 \mathcal{G}^a_{ij,mn}(t) &= \mathcal{G}^{r*}_{mn,ij}(-t)
 \nonumber \\
 & = i \theta(-t) \ll \hat X_{mn} | \mathcal{U}_{eff}(-t) | \hat X_{ij} \gg 
\end{align}

Note, that definitions (\ref{GrNew}) and (\ref{GaNew}) lead to the
usual Hermitian-type connection (\ref{GaGrNew}) between retarded and
advance Green functions $\mathcal{G}^a=[\mathcal{G}^r]^\dagger$. 
An alternative definition
\begin{align}
 \label{GrNewalt}
 \mathcal{G}^r_{ij,mn}(t) &\equiv -i\theta(t) 
 \ll \hat X_{ji} \hat{I}_{K} | e^{-i \mathcal{L} t} | \hat X_{mn} \hat \rho_{K}^{eq} \gg
 \\ \nonumber
 & = -i\theta(t) \ll \hat X_{ji}| \mathcal{U}_{eff}(t) | \hat X_{mn} \gg \\
 \label{GaNewalt}
 \mathcal{G}^a_{ij,mn}(t) &\equiv i \theta(-t)
 \ll \hat X_{nm} \hat{I}_{K} | e^{i \mathcal{L} t} | \hat X_{ij} \hat \rho_{K}^{eq}  \gg 
 \\ \nonumber
 & = i \theta(-t) \ll \hat X_{ji} | \mathcal{U}_{eff}^{\dagger}(-t) | \hat X_{mn} \gg  \nonumber
\end{align}
would lead to Liouvillian conjugation\cite{BenReuven}
$\mathcal{G}^a=[\mathcal{G}^r]^\times$ or
\begin{equation}
 \mathcal{G}^a_{ij,mn}(t) = \mathcal{G}^{r*}_{nm,ji}(-t)
\end{equation}

\section{\label{appC}Expression for $\mathcal{L}_{eff}$}

We start from (\ref{Gr}) and (\ref{Ga}) and use free propagator 
in place of effective one. This leads to
\begin{align}
 \label{Gr_free}
 \mathcal{G}^{(0)\, r}_{ij,mn}(t) &= 
 -i\theta(t) \ll \hat X_{ji} | e^{-i\mathcal{L}_M t} | \hat X_{nm} \gg
 \\
 & \equiv -i \theta(t) <j| e^{-i\hat H_M t} |n>\, 
                       <m| e^{ i\hat H_M t} |i> 
 \nonumber \\
 \label{Ga_free}
 \mathcal{G}^{(0)\, a}_{ij,mn}(t) &=
 i\theta(-t) \ll \hat X_{ji} | e^{-i\mathcal{L}^\dagger_M t} | \hat X_{nm} \gg
 \\
 & \equiv i\theta(-t) <j| e^{-i\hat H_M t} |n>\,
                      <m| e^{ i\hat H_m t} |i>
 \nonumber
\end{align}
Substituting (\ref{Gr_free}) and (\ref{Ga_free}) into (\ref{gen_QME})
and using standard Markov approximation
\begin{equation}
 \rho_{ab}(t_1) \approx \sum_{c,d} \ll ab | e^{i\mathcal{L}_M (t-t_1)} | cd \gg
 \rho_{cd}(t)
\end{equation}
one gets the (Markovian) Redfield quantum master equation
\begin{equation}
\label{Redfield}
  \frac{d \rho_{ab}(t)}{dt} = 
  -i \sum_{c,d} \ll ab | \mathcal{L}_{eff} | cd \gg \rho_{cd}(t) ,
  \nonumber
\end{equation}
where the generator for our model takes the form
\begin{widetext}
\begin{align}
 \label{Leff}
 &-i\mathcal{L}^{eff}_{(a;b),(c;d)}
 = i\mathcal{L}^{eff\,\dagger}_{(b;a),(d;c)}
 =-i\left\{
  \delta_{N_a,N_c}\delta_{N_b,N_d}\left[
  H^{(N_a)}_{s_a,s}\delta_{s_b,s_d} - \delta_{s_a,s_c}H^{(N_b)}_{s_d,s_b}
 \right]
 -\frac{1}{2}\sum_{i,j}\sum_{p,r}
 \right. \nonumber \\
 & \left[
 \delta_{N_a+1,N_c}\delta_{N_b+1,N_d}(-1)^{N_a-N_b}\times
 \right. \nonumber \\
 & \qquad \left(
  U^{(N_a+1)}_{ri}{\overset{*}{U}}{}^{(N_a+1)}_{s_ci}
  U^{(N_a)}_{s_aj}{\overset{*}{U}}{}^{(N_a)}_{pj}
  \Sigma^{>}_{(N_b,s_b;N_b+1,s_d),(N_a,p;N_a+1,r)}(E_i^{(N_a+1)}-E_j^{(N_a)})
 \right.
 \nonumber \\
 & \left. \qquad
 +U^{(N_b+1)}_{s_di}{\overset{*}{U}}{}^{(N_b+1)}_{ri}
  U^{(N_b)}_{pj}{\overset{*}{U}}{}^{(N_b)}_{s_bj}
  \Sigma^{>}_{(N_b,p;N_b+1,r),(N_a,s_a;N_a+1,s_c)}(E^{(N_b+1)}_i-E^{(N_b)}_j)
 \right)
 \nonumber \\
 &-\delta_{N_a-1,N_c}\delta_{N_b-1,N_d}(-1)^{N_a-N_b}\times
 \nonumber \\
 & \qquad \left(
  U^{(N_a)}_{s_ai}{\overset{*}{U}}{}^{(N_a)}_{ri}
  U^{(N_a-1)}_{pj}{\overset{*}{U}}{}^{(N_a-1)}_{s_cj}
  \Sigma^{<}_{(N_a-1,p;N_a,r),(N_b-1,s_d;N_b,s_b)}(E^{(N_a)}_i-E^{(N_a-1)}_j)
 \right.
 \nonumber \\
 & \left. \qquad
 +U^{(N_b)}_{ri}{\overset{*}{U}}{}^{(N_b)}_{s_bi}
  U^{(N_b-1)}_{s_dj}{\overset{*}{U}}{}^{(N_b-1)}_{pj}
  \Sigma^{<}_{(N_a-1,s_c;N_a,s_a),(N_b-1,p;N_b,r)}(E^{(N_b)}_i-E^{(N_b-1)}_j)
 \right)
 \\
 &+\delta_{N_a,N_c}\delta_{N_b,N_d}\delta_{s_a,s_c}
 \nonumber \\
 & \ \ \sum_s \left(
  U^{(N_b+1)}_{ri}{\overset{*}{U}}{}^{(N_b+1)}_{si}
  U^{(N_b)}_{s_dj}{\overset{*}{U}}{}^{(N_b)}_{pj}
  \Sigma^{<}_{(N_b,s_b;N_b+1,s),(N_b,p;N_b+1,r)}(E^{(N_b+1)}_i-E^{(N_b)}_j)
 \right.
 \nonumber \\
 & \left. \qquad
 -U^{(N_b)}_{s_di}{\overset{*}{U}}{}^{(N_b)}_{ri}
  U^{(N_b-1)}_{pj}{\overset{*}{U}}{}^{(N_b-1)}_{sj}
  \Sigma^{>}_{(N_b-1,p;N_b,r),(N_b-1,s;N_b,s_b)}(E^{(N_b)}_i-E^{(N_b-1)}_j)
 \right)
 \nonumber \\
 &+\delta_{N_a,N_c}\delta_{N_b,N_d}\delta_{s_b,s_d}
 \nonumber \\
 & \ \ \sum_s \left(
  U^{(N_a+1)}_{si}{\overset{*}{U}}{}^{(N_a+1)}_{ri}
  U^{(N_a)}_{pj}{\overset{*}{U}}{}^{(N_a)}_{s_cj}
  \Sigma^{<}_{(N_a,p;N_a+1,r),(N_a,s_a;N_a+1,s)}(E^{(N_a+1)}_i-E^{(N_a)}_j)
 \right.
 \nonumber \\
 & \left. \left. \left.\qquad
 -U^{(N_a)}_{ri}{\overset{*}{U}}{}^{(N_a)}_{s_ci}
  U^{(N_a-1)}_{sj}{\overset{*}{U}}{}^{(N_a-1)}_{pj}
  \Sigma^{>}_{(N_a-1,s;N_a,s_a),(N_a-1,p;N_a,r)}(E^{(N_a)}_i-E^{(N_a-1)}_j)
 \right)
 \right]\right\}
 \nonumber 
\end{align}
\end{widetext}
where $\mathbf{U}^{(N)}$ are unitary transformations diagonalizing 
charge blocks $\mathbf{H}_M^{(N)}$ of the molecular Hamiltonian (\ref{HM}),
and $E^{(N)}_i$ are corresponding eigenvalues.


\end{document}